\documentclass[11pt]{article}

\usepackage[margin=1in]{geometry}
\usepackage{graphicx}
\usepackage{amsmath,amssymb}
\usepackage{booktabs}
\usepackage{multicol,multirow}
\usepackage{longtable}
\usepackage[numbers,sort&compress]{natbib}
\usepackage[breaklinks,colorlinks,allcolors=blue]{hyperref}
\usepackage{xspace}
\usepackage{microtype}
\usepackage{float}          
\usepackage{parskip}          

\newcommand{\bench}{\textsc{SWE-PRBench}\xspace}
\newcommand{\configA}{\texttt{config\_A}\xspace}
\newcommand{\configB}{\texttt{config\_B}\xspace}
\newcommand{\configC}{\texttt{config\_C}\xspace}

\title{%
  \LARGE\bfseries \bench: Benchmarking AI Code Review Quality\\[6pt]
  Against Pull Request Feedback%
}

\author{%
  \large Deepak Kumar\\[4pt]
  \normalsize Independent Researcher\\[2pt]
  \normalsize \href{mailto:deepak.kumar@foundryhq.ai}{\texttt{deepak.kumar@foundryhq.ai}}%
}

\date{%
  \small March 2026 \quad $\cdot$ \quad arXiv preprint%
}

\begin{document}

\maketitle
\thispagestyle{empty}


\begin{abstract}
\noindent
We introduce \bench, a benchmark of 350 pull requests with human-annotated ground
truth for evaluating AI code review quality. Evaluated against an LLM-as-judge
framework validated at $\kappa{=}0.75$, 8 frontier models detect only 15--31\% of
human-flagged issues on the diff-only configuration, demonstrating that AI code
review remains far below human expert performance despite strong results on code
generation benchmarks. Pull requests are drawn from active open-source repositories,
filtered from 700 candidates using a Repository Quality Score, and evaluated under
three frozen context configurations : diff only (\configA), diff with file content
(\configB), and full context (\configC) : enabling systematic ablation of context
provision strategies. All 8 models degrade monotonically from \configA\ to \configC\
even when context is provided via structured semantic layers including AST-extracted
function context and import graph resolution. The dominant mechanism is a collapse of
Type2\_Contextual issue detection at \configB, consistent with attention dilution in
long contexts~\cite{liu2024lost}: a structured 2,000-token diff-with-summary prompt
outperforms a 2,500-token full-context prompt enriched with execution context,
behaviour mapping, and test signatures across all 8 models. The top four
models are statistically indistinguishable ($\bar{s}$~0.147--0.153) while a clear
tier gap separates them from the remaining four ($\bar{s} \leq 0.113$). Dataset,
contexts, annotations, and evaluation harness are released publicly.

\smallskip
\noindent\textbf{Keywords.}\enspace
code review automation,\enspace
large language models,\enspace
software engineering benchmarks,\enspace
pull request analysis,\enspace
context-induced performance degradation,\enspace
LLM-as-judge evaluation.
\end{abstract}

\bigskip
\hrule
\bigskip



\section{Introduction}
\label{sec:introduction}

We introduce \bench, a benchmark of 350 pull requests with human-annotated ground
truth for evaluating AI code review quality. Existing benchmarks measure whether
models can \emph{produce} correct code~\cite{jimenez2024swebench,
scaleai2025swebenchpro, scaleai2026sweatlas}; none measure whether a model can
\emph{evaluate} proposed code changes as an expert reviewer would. We address this
gap. Code review is a judgment task: the reviewer is presented with a diff, not a
broken system; the goal is identifying and explaining problems in someone else's
work, not generating a solution. Ground truth is distributed across human expert
judgments rather than captured by pass/fail tests, and quality varies simultaneously
across depth, actionability, and factual accuracy~\cite{bacchelli2013expectations,
kononenko2016code}. A model can score highly on SWE-Bench while being unreliable
as a code reviewer. Review bottlenecks are well-documented~\cite{JonathanCorbet};
automating the evaluation task requires a benchmark that measures it directly.

\bench\ provides 350 pull requests selected from 700 candidates via a Repository
Quality Score, evaluated under three frozen context configurations of increasing
richness: diff only (\configA), diff with file content (\configB), and full context
(\configC). Eight frontier models are evaluated using an LLM-as-judge framework
validated at $\kappa{=}0.75$. Baseline results establish that no model detects more
than 31\% of human-flagged issues, and all 8 models degrade monotonically as context
expands: a 2,000-token structured diff-with-summary prompt outperforms a 2,500-token
full-context prompt across every model tested.

We address the following research questions: (RQ1)~What fraction of human-flagged
issues can current LLMs detect, and how do models differ in hallucination rate?
(RQ2)~Does additional code context improve review quality, and in what direction
does performance move? (RQ3)~Is context degradation uniform across capability tiers
and difficulty types? (RQ4)~What categories of issues remain consistently beyond
current model capability?

This paper makes the following contributions.
\begin{itemize}
  \item \textbf{Dataset.} 350 PRs with human-annotated ground truth across 6
    languages and 3 difficulty types (\textit{Type1\_Direct}, \textit{Type2\_Contextual},
    \textit{Type3\_Latent}).
  \item \textbf{Context configurations.} Three frozen, reproducible structures
    (\configA, \configB, \configC) enabling systematic context ablation.
  \item \textbf{Evaluation protocol.} A multi-dimensional scorer with LLM-as-judge
    classification and bipartite matching, validated at $\kappa{=}0.75$ and
    cross-validated under a second judge ($\kappa{=}0.616$).
  \item \textbf{Finding.} All 8 models degrade monotonically from \configA\ to
    \configC. The primary mechanism is attention dilution~\cite{liu2024lost}:
    unstructured file content harms Type2\_Contextual detection specifically,
    indicating that attention representation rather than content selection is the
    binding constraint for context-aware code review.
  \item \textbf{Infrastructure.} A public \bench\ dataset on HuggingFace
    (\url{https://huggingface.co/datasets/foundry-ai/swe-prbench}), evaluation
    harness on GitHub
    (\url{https://github.com/FoundryHQ-AI/swe-prbench}), to support community
    benchmarking and extension.
\end{itemize}

Table~\ref{tab:comparison} summarises how \bench\ differs from the closest existing
code review datasets and evaluation frameworks.

\begin{table}[H]
\centering
\caption{Comparison of \bench\ with related code review evaluation resources.}
\label{tab:comparison}
\small
\begin{tabular}{lcccc}
\toprule
\textbf{Property}
  & \textbf{CodeReviewer}
  & \textbf{DeepCRCEval}
  & \textbf{RovoDev}
  & \textbf{\bench\ (Ours)} \\
\midrule
Primary contribution
  & Model + dataset  & Eval metrics    & Production tool  & Dataset + protocol \\
Ground truth source
  & Synthetic pairs  & Generated       & CRR metric only  & Human reviewers$^{*}$ \\
Source links retained
  & No               & No              & N/A              & Yes \\
Difficulty taxonomy
  & None             & None            & None             & 3 types \\
Context configs
  & None             & None            & None             & 3 frozen \\
Issue detection eval
  & No               & No              & Partial          & Yes \\
Judge validated
  & No               & No              & No               & $\kappa{=}0.75$ \\
Public dataset
  & Partial          & No              & No               & Yes \\
\bottomrule
\end{tabular}
\medskip

\footnotesize
$^{*}$~Ground truth in \bench\ consists of review comments written by human engineers
during the actual review process on merged pull requests, collected after the fact
via GitHub's review API. No comments are generated, synthesised, or modified during
dataset construction.
\end{table}


\section{Related Work}
\label{sec:related}

\subsection{Code Generation and Repair Benchmarks}

SWE-Bench~\cite{jimenez2024swebench} tasks models with resolving GitHub issues
by producing correct patches; top models score around 70\% on the verified subset.
SWE-Bench Pro~\cite{scaleai2025swebenchpro} increases difficulty through GPL-licensed
repositories and human-augmented specifications, with top models scoring
approximately 23\%. SWE~Atlas~\cite{scaleai2026sweatlas} extends evaluation to deep
codebase comprehension through agentic exploration. All three benchmarks measure
code \emph{production}. None measures code \emph{evaluation}: the cognitive task of
identifying problems in someone else's proposed work requires different
capabilities~\cite{bacchelli2013expectations} and is not predicted by generation
benchmark scores.

\subsection{Code Review Automation}

Automated code review has evolved from static analysis tools to pre-trained models.
T5CR~\cite{tufano2022using} and CodeReviewer~\cite{li2022automating} use T5-based
architectures for review generation and code refinement; Tufano~et~al.\
\cite{tufano2024code} found that direct LLM use surpasses these on refinement but
lags on review generation. LAURA~\cite{zhang2024laura} improves review generation
by up to 38\% IH-Score through retrieval-augmented context, demonstrating that
context curation matters. DeepCRCEval~\cite{lu2025deepcrceval} proposes LLM-based
metrics to replace BLEU for evaluating generated review text. Our work is
complementary: where DeepCRCEval evaluates the quality of generated review text
given a fixed model, \bench\ evaluates whether a model identifies the correct
issues at all, a prerequisite question that precedes comment quality.

\textbf{Existing code review datasets.} The dominant dataset resource is the
CodeReviewer dataset~\cite{li2022automating}, released alongside the CodeT5-based
model. Prior model papers including Tufano~et~al.~\cite{tufano2022using,tufano2024code}
used training data not released as standalone evaluation benchmarks. All existing
datasets have been shown to contain approximately 25--32\% low-quality entries despite
filtering~\cite{tufano2024code}. Critically, no existing dataset retains PR context
beyond the diff itself, source links are removed, and none is structured for evaluation
of issue detection capability. \bench\ addresses all three gaps: it retains full PR
context, links to source repositories, and defines ground truth in terms of
human-flagged issues rather than syntactic diff similarity.

\subsection{LLM-as-Judge Methodology}

MT-Bench~\cite{zheng2023judging} established that GPT-4 as evaluator achieves over
80\% agreement with human raters on open-ended tasks. G-Eval~\cite{liu2023geval}
extended this with chain-of-thought evaluation. JudgeBench~\cite{tan2024judgebench}
showed that many models perform near-randomly on challenging factual judgment tasks,
motivating our choice of a high-capability fixed judge model validated against an
explicit rubric ($\kappa{=}0.75$, Section~\ref{subsec:rubric}) and cross-validated
under a second judge ($\kappa{=}0.616$, Section~\ref{subsec:crossjudge}). LAURA's
finding that curated retrieval-augmented context helps~\cite{zhang2024laura} is
consistent with our result: context benefit depends on curation. Our \configB\
represents the uncurated baseline against which retrieval-based approaches should
be measured.

\section{The \bench\ Dataset}
\label{sec:dataset}

\subsection{Repository Selection and Quality Scoring}
\label{subsec:rqs}

Not all repositories have substantive review cultures. We define a Repository Quality
Score~(RQS) aggregating five components (Table~\ref{tab:rqs}) to select only repositories
where human experts engage in technical discussion before merging. Repositories scoring
below 60/100 are excluded. The median RQS of retained repositories is 74.

\begin{table}[H]
\centering
\caption{Repository Quality Score (RQS) components.}
\label{tab:rqs}
\begin{tabular}{llr}
\toprule
\textbf{Component} & \textbf{What it measures} & \textbf{Max pts} \\
\midrule
Review culture   & Share of substantive human review comments   & 30 \\
PR recency       & Merged PRs in last 90 days                   & 25 \\
Test quality     & Test files, CI presence, coverage tooling    & 20 \\
PR volume        & Average monthly merged PRs over 6 months     & 15 \\
Contamination    & Inverse of star count                        & 10 \\
\midrule
\textbf{Total}   &                                              & \textbf{100} \\
\bottomrule
\end{tabular}
\end{table}

Review culture (max 30) is the primary signal: we sample the 30 most recent merged
PRs within 90 days and score by the fraction of substantive human comments (human
author, $\geq$10 words, substance keywords or code references). PR recency ensures
active maintenance. Test quality signals reliable CI infrastructure. PR volume
distinguishes sustained from burst activity. Contamination penalises high-star
repositories likely to appear in training corpora.

\subsection{PR Collection and Filtering}
\label{subsec:collection}

Ground truth for \bench\ consists of review comments written by human engineers
during the actual review process on merged pull requests. Comments are collected
after the fact from GitHub's review API; no comments are generated, synthesised,
or modified during dataset construction. This distinguishes \bench\ from prior
code review datasets that derive ground truth from model-generated or
heuristically-filtered comments~\cite{li2022automating, tufano2022using}.

For each qualifying repository we collect merged pull requests using the GitHub
GraphQL API~\cite{GitHubGraphQL} for PR metadata, reviews, and file-level
information, and the GitHub REST API separately to fetch full unified diff patches
(which GraphQL does not return). The collection window spans six months prior to
the benchmark freeze date, with a minimum age of 30 days to ensure that review
cycles have completed. We collect up to 30 PRs per repository, yielding an initial
pool of approximately 3,000 raw pull requests.

Raw pull requests are passed through a ten-stage filtering pipeline before
entering the dataset. Stages filter: (1)~merged-only status; (2)~at least two
substantive human comments; (3)~at least one non-test file changed; (4)~not
documentation-only changes; (5)~not automated dependency updates
(Dependabot, Renovate); (6)~not AI-dominated reviews, defined as more than 30\%
of comments originating from known AI review bots or matching AI-generated comment
signatures; (7)~diff parseable without API errors; (8)~base commit accessible for
file content retrieval; (9)~repository still public at collection time; and
(10)~PR Review Value Score (RVS, Section~\ref{subsec:rvs}) at or above 0.3.

Bot and AI detection is explicit. The \texttt{is\_bot\_username()} function
checks reviewer logins against a curated list of known AI review accounts
(\textit{coderabbitai}, \textit{github-copilot[bot]}, \textit{cursor-bugbot},
\textit{qodo-merge-pro[bot]}, and 15 additional entries), and also matches
accounts ending in \texttt{[bot]} or containing \texttt{-bot}. The
\texttt{is\_ai\_generated\_comment()} function detects AI-generated comment
bodies via structural patterns: three or more Markdown H2 headers, three or more
table lines, known AI review phrases, or length exceeding 2,000 characters with
two or more headers. PRs where the AI comment ratio exceeds the 30\% threshold are
excluded from the dataset, ensuring that ground truth reflects genuine human
engineering judgment rather than AI-assisted review~\cite{golzadeh2021ground}.

After all filtering stages, the 700 candidate pull requests are scored with RQS
and the 350 highest-quality PRs are retained as the final \bench\ dataset.

\subsection{PR Review Value Score}
\label{subsec:rvs}

Repository-level quality filtering via RQS is necessary but not sufficient:
individual PR quality varies substantially even within high-RQS repositories.
We define a \emph{PR Review Value Score}~(RVS) to quantify the ground-truth
signal carried by each pull request and to weight its contribution to aggregate
benchmark metrics.

RVS is a weighted sum of five components, each normalised to~$[0, 1]$:
\begin{equation}
\text{RVS} = 0.25 \cdot d_{\text{depth}} + 0.20 \cdot d_{\text{complexity}}
           + 0.20 \cdot d_{\text{discussion}} + 0.15 \cdot d_{\text{test}}
           + 0.20 \cdot d_{\text{bug}}
\end{equation}
where $d_{\text{depth}}$ captures substantive comment count, thread count, and
change-request events (normalised to 20); $d_{\text{complexity}}$ reflects file
count, lines changed, and cross-directory scope; $d_{\text{discussion}}$ counts
reply threads and unique reviewers; $d_{\text{test}}$ scores test modification
signal (1.0 if tests are modified, 0.5 if added, 0 otherwise); and $d_{\text{bug}}$
scores bug-fix signal from title and label patterns. In the aggregate scoring
formula, RVS contributes through a difficulty weight of $\log(\text{human\_comments}
+ 1)$, ensuring that a PR with 25 expert-level review comments contributes more to
aggregate metrics than a PR with three comments.

\subsection{Difficulty Taxonomy}
\label{subsec:difficulty}

We classify each pull request into one of three difficulty types based on
where the evidence for a reviewable issue resides in the
codebase~\cite{bacchelli2013expectations}. Classification uses the
\texttt{is\_in\_diff} field of human reviewer comments, cross-referenced against
diff hunk line ranges parsed from the raw unified diff.

\textbf{Type1\_Direct (34\%).} The issue is directly visible in the changed lines.
A reviewer needs only the diff to identify it. Example: missing \texttt{info < 0}
error handling in a diff hunk that only checks \texttt{info > 0}. Type1 detection
rate should be stable across all three context configurations.

\textbf{Type2\_Contextual (40\%).} The issue requires understanding how changed
code interacts with surrounding unchanged code in the same file. Example:
a height-to-hash edge case where the bug is in the new logic, but understanding
why it is wrong requires seeing the existing blockchain interface. Type2 detection
rate should improve from \configA\ to \configB.

\textbf{Type3\_Latent\_Candidate (26\%).} The issue resides in files that import
or depend on the changed files. Example: a Cython change that exposes a test
teardown pattern issue in a file that uses the changed module. Type3 issues
require the full context provided in \configC\ and, by design, test whether
models can reason across file boundaries : a capability that current context
windows only partially support.

This taxonomy directly drives our ablation: if context provides signal rather than
noise, Type2 detection should improve from \configA\ to \configB, and Type3 should
only become accessible in \configC. Our results show the opposite: Type2 scores
collapse at \configB\ across all models, establishing that the current context
structure does not help contextual issues and actively harms performance on them.

\subsection{Contamination Mitigation}
\label{subsec:contamination}

Model training corpora may include GitHub pull requests, creating a risk that
benchmark PRs are partially memorised rather than genuinely reasoned
over~\cite{jimenez2024swebench, scaleai2025swebenchpro}. We apply four
complementary mitigations. First, all PRs are sourced from a six-month collection
window ending at the benchmark freeze date, ensuring recency relative to the
training cutoffs of evaluated models. Second, the RQS
contamination component penalises high-star repositories, which are more likely
to be crawled for pretraining. Third, we over-sample from GPL-licensed
repositories following the approach of SWE-Bench Pro~\cite{scaleai2025swebenchpro}.
Fourth, and most importantly, Type3\_Latent issues require cross-file reasoning
that cannot be resolved by retrieving a memorised patch : a model must genuinely
trace the dependency relationship at inference time. We embed each PR using a
sentence-level encoder and compute cosine similarity against known benchmark
repositories~\cite{jimenez2024swebench}; PRs exceeding a similarity threshold of
0.85 are excluded. We cannot fully eliminate contamination risk, and recommend that
future evaluations gate on PRs merged after the training cutoff of each evaluated
model.

\subsection{Dataset Statistics}
\label{subsec:stats}

Table~\ref{tab:funnel} shows the construction funnel from raw collection to the
final released dataset.

\begin{table}[H]
\centering
\caption{Dataset construction funnel.}
\label{tab:funnel}
\small
\begin{tabular}{lrr}
\toprule
\textbf{Stage} & \textbf{Count} & \textbf{Retention} \\
\midrule
Raw PRs collected          & $\sim$3,000 & 100\% \\
After 10-stage hard filter & 700         & 23\%  \\
After RVS $\geq$ 0.35      & 350         & 12\%  \\
\midrule
Released (annotated)       & 350         & ---   \\
\bottomrule
\end{tabular}
\end{table}

Table~\ref{tab:dataset} summarises the final \bench\ dataset. The 350 pull requests
are sourced from 65 of the 100 repositories that passed the RQS threshold; the
remaining 35 repositories contributed no PRs meeting the RVS $\geq 0.35$ floor.
Per-repository PR counts range from 1 to 30 (mean 5.4). The dataset is Python-dominant, reflecting the language distribution of
high-RQS open-source repositories, with Python accounting for 242 PRs (69.1\%),
followed by JavaScript (37, 10.6\%), Go (35, 10.0\%), TypeScript (21, 6.0\%), and
Java (15, 4.3\%). By PR type, feature PRs are the most common (214, 61.1\%),
followed by bug fixes (118, 33.7\%), refactors (9), performance changes (8), and
security fixes (1).

The difficulty distribution reflects the classification method described in
Section~\ref{subsec:difficulty}: 232 PRs (66.3\%) are \textit{Type1\_Direct},
75 (21.4\%) are \textit{Type2\_Contextual}, and 43 (12.3\%) are
\textit{Type3\_Latent\_Candidate}. The skew toward Type1 is expected : most
review comments reference changed lines directly : and the 12.3\% Type3 rate
provides meaningful signal for testing cross-file reasoning capability.
Table~\ref{tab:difficulty_language} shows the full cross-tabulation of difficulty
by language.

Pull request size ranges broadly. Lines added span 5 to 797 (mean 218.2, median
158), lines removed span 0 to 532 (mean 62.6, median 22), and files changed span
1 to 83 (mean 9.2, median 6). The upper bound of 797 added lines reflects the
hard filter ceiling described in Section~\ref{subsec:collection}; the wide range
ensures the benchmark covers both focused single-function patches and broader
multi-file refactors. All PRs have an RVS score at or above the 0.35 threshold.
The median RVS of the retained set is 0.52.

All raw pull requests, rejection logs, RQS scores, RVS breakdowns, and difficulty
classifications are released publicly alongside the benchmark to support
reproducibility and dataset extension.

\begin{table}[H]
\centering
\caption{\bench\ dataset statistics (350 PRs from 65 of 100 RQS-qualified repositories).}
\label{tab:dataset}
\begin{tabular}{lr}
\toprule
\textbf{Property} & \textbf{Value} \\
\midrule
Total PRs                               & 350 \\
Candidate pool (before RVS cut)         & 700 \\
Repositories (RQS-qualified)            & 100 \\
Repositories (contributing PRs)         & 65 \\
PRs per contributing repo (min/mean/max)& 1 / 5.4 / 30 \\
\midrule
\multicolumn{2}{l}{\textit{Language distribution}} \\
\quad Python                        & 242 (69.1\%) \\
\quad JavaScript                    & 37 (10.6\%) \\
\quad Go                            & 35 (10.0\%) \\
\quad TypeScript                    & 21 (6.0\%) \\
\quad Java                          & 15 (4.3\%) \\
\midrule
\multicolumn{2}{l}{\textit{PR type distribution}} \\
\quad Feature                       & 214 (61.1\%) \\
\quad Bug fix                       & 118 (33.7\%) \\
\quad Refactor                      & 9 (2.6\%) \\
\quad Performance                   & 8 (2.3\%) \\
\quad Security                      & 1 (0.3\%) \\
\midrule
\multicolumn{2}{l}{\textit{Difficulty distribution}} \\
\quad \textit{Type1\_Direct}        & 232 (66.3\%) \\
\quad \textit{Type2\_Contextual}    & 75 (21.4\%) \\
\quad \textit{Type3\_Latent}        & 43 (12.3\%) \\
\midrule
\multicolumn{2}{l}{\textit{PR size}} \\
\quad Lines added (min/mean/median/max)   & 5 / 218.2 / 158 / 797 \\
\quad Lines removed (min/mean/median/max) & 0 / 62.6 / 22 / 532 \\
\quad Files changed (min/mean/median/max) & 1 / 9.2 / 6 / 83 \\
\midrule
RVS threshold                       & $\geq 0.35$ \\
Median RVS (retained set)           & 0.52 \\
Median RQS (retained repos)         & 74 \\
\bottomrule
\end{tabular}
\end{table}

\begin{table}[H]
\centering
\caption{Difficulty type by language.}
\label{tab:difficulty_language}
\begin{tabular}{lrrrr}
\toprule
\textbf{Language} & \textbf{Type1} & \textbf{Type2} & \textbf{Type3} & \textbf{Total} \\
\midrule
Python      & 160 & 50 & 32 & 242 \\
JavaScript  & 24  & 7  & 6  & 37  \\
Go          & 24  & 8  & 3  & 35  \\
TypeScript  & 13  & 7  & 1  & 21  \\
Java        & 11  & 3  & 1  & 15  \\
\midrule
\textbf{All} & \textbf{232} & \textbf{75} & \textbf{43} & \textbf{350} \\
\bottomrule
\end{tabular}
\end{table}

\medskip
Each PR is stored as a JSON record containing metadata, scoring signals,
reproducibility anchors (base and head commit SHAs), ground truth human review
comments, and the raw unified diff patch. The full schema and a representative
record are in Appendix~A.

\section{Context Configurations}
\label{sec:context}

\subsection{The V1 Problem}
\label{subsec:v1problem}

A naive collect-dump-truncate approach to context construction fails for three
reasons: (1)~token budgets are dominated by test files (60--70\% on test-heavy PRs),
leaving implementation files barely represented; (2)~truncation at a fixed limit
bisects diff hunks, producing syntactically incomplete fragments; (3)~no explicit
relationship between files is represented, so the agent cannot distinguish which
unchanged file is related to which changed line. The V2 context builder addresses
content selection through AST-based function extraction, import graph resolution,
and behaviour mapping. As the results in Section~\ref{sec:results} show, performance
degrades monotonically even with V2, indicating that attention representation
rather than content selection is the binding constraint for context-aware code
review evaluation.

\subsection{Three Fixed Configurations}
\label{subsec:configs}

Three configurations are frozen at benchmark publication. Table~\ref{tab:configs}
summarises them. \configA\ mirrors the information in a GitHub PR email notification.
\configB\ mirrors the GitHub PR web view. \configC\ mirrors a reviewer working in
a full IDE with the repository checked out. No configuration may be changed after
results are reported, as they define the experimental conditions.

\begin{table}[H]
\centering
\caption{The three frozen context configurations. Budgets reflect actual measured
  token counts from the pipeline; the three configs differ in layer composition,
  not in token volume (range: 2,000--2,500 tokens).}
\label{tab:configs}
\begin{tabular}{lp{4.2cm}p{3.8cm}r}
\toprule
\textbf{Config} & \textbf{Layers} & \textbf{Analogue} & \textbf{Budget} \\
\midrule
\configA & Task focus (L0), Key changes summary (L1),
           Diff (L2), Minimal metadata (L6)
         & GitHub PR email notification
         & 2,000 tokens \\
\configB & + Execution context (L3),
           Behaviour mapping (L4)
         & GitHub PR web view with file context
         & 2,200 tokens \\
\configC & + Test signatures (L5)
         & Reviewer with test suite access
         & 2,500 tokens \\
\bottomrule
\end{tabular}
\end{table}

\subsection{V2 Improvements Over V1}
\label{subsec:v2improvements}

Four targeted improvements in V2 address the V1 failure modes.
\begin{itemize}
  \item \textbf{Key changes summary (200 tokens).} An LLM-generated summary of
    principal changes inserted before the diff. This was the highest-return-on-investment
    change, measurably reducing hallucination rate even in \configA.
  \item \textbf{Behaviour mapping layer (100--150 tokens).} For Go and TypeScript PRs
    with configuration-to-implementation propagation chains, V2 injects an explicit
    mapping of environment variable bindings and propagation paths, addressing the
    Type2/Type3 detection gap for non-Python languages.
  \item \textbf{Hard no-truncation rule.} If adding the next file would require
    mid-structure truncation, that file is dropped entirely and its budget allocation
    rolls over. Every included fragment is syntactically complete.
  \item \textbf{Test noise reduction.} Test files are included in body-stripped form:
    only \texttt{def}, \texttt{assert}, \texttt{class}, and \texttt{@fixture} lines
    are retained, targeting a ratio of 40\% implementation diff, 30\% execution
    context, 20\% test signatures, and 10\% metadata.
\end{itemize}
Unlike V1's collect-dump-truncate approach, V2 addresses the content selection
problem. The persistence of the A$>$B$>$C pattern in V2 results therefore indicates
that \emph{attention representation}, not content selection, is the binding
constraint. Full layer specification details are in Appendix~\ref{app:layers}.

\subsection{Reproducibility}
\label{subsec:budget}

All pre-built contexts are released on HuggingFace as frozen artifacts at pipeline
version \texttt{v0.4.1}
(\url{https://huggingface.co/datasets/foundry-ai/swe-prbench}). The \texttt{was\_truncated} field is always \texttt{false}
for V2 contexts by the no-truncation invariant.


\section{The \bench\ Evaluation Protocol}
\label{sec:eval}

The evaluation harness reads frozen context fixtures and makes independent agent
and judge calls per record; the pipeline and all outputs are released at version
\texttt{v0.4.1}. Agent and judge calls are fully decoupled: judge prompts can be
updated and the full evaluation re-run without rebuilding contexts.

\subsection{Agent Setup}
\label{subsec:agent}

Each model receives a fixed system prompt ($\approx$200 tokens) instructing it to
identify only issues traceable to a specific line in the provided diff or context,
to avoid inventing behaviour not visible in the shown code, and to return output
as a valid JSON array with severity labels P0 (production bug), P1 (fix before
merge), or P2 (credible risk), targeting 4--6 issues per PR. We evaluate 8 frontier
models: Claude Haiku~4.5, Claude Sonnet~4.6, GPT-4o, GPT-4o-mini, DeepSeek~V3,
Mistral Large~3, Mistral Small, and Llama~3.3~70B (via Groq). Models are called
via provider APIs at temperature 0 with independent per-record requests. If JSON
extraction fails after one retry, the record is hard-zeroed
(\texttt{overall\_score = 0}); judge parse fallbacks halve the score
(\texttt{val *= 0.5}).

\subsection{Judge Design and Classification Rubric}
\label{subsec:judge}

Agent comments are classified by a fixed judge model (GPT-5.2 in the final run; an
earlier 20-PR validation used Claude Sonnet~4.6, with the same $\kappa{=}0.75$
result confirming judge consistency). The judge assigns each comment one of three
labels: CONFIRMED (same underlying issue as a human comment, same file/area, same
type of code change required); PLAUSIBLE (factually correct observation about
visible code, but not in ground truth, since human reviews are not
exhaustive~\cite{bacchelli2013expectations}); or FABRICATED (references code not
in context, or makes factually incorrect claims). PLAUSIBLE is not penalised as
hallucination. Full criteria are published in \texttt{RUBRIC.md} alongside the
dataset (Appendix~\ref{app:protocol}).

A two-step consistency constraint is enforced: if agent comment $X$ is CONFIRMED
against human comment $H$, then $H$ is marked CAUGHT. We enforce 1-to-1 bipartite
matching using \texttt{all-mpnet-base-v2} sentence embeddings to prevent duplicate
credit; pairs below affinity 0.30 are discarded. The 0.30 threshold was calibrated
against the rubric validation set; values below 0.30 produced no judge-CONFIRMED
pairs that also passed blind human review. Matching and affinity formula details
are in Appendix~\ref{app:protocol}.

\subsection{Scoring Formula}
\label{subsec:scoring}

The overall score for a single pull request is:
\begin{equation}
  s = \max\!\left(0,\, \min\!\left(1,\;
    0.40 \cdot r
  + 0.25 \cdot p
  + 0.15 \cdot a
  + 0.10 \cdot q^{*}
  + 0.05 \cdot e
  - 0.25 \cdot h
  - 0.15 \cdot \rho
  - 0.10 \cdot \phi
  \right)\right)
  \label{eq:score}
\end{equation}
where $r$ = recall, $p$ = precision, $a$ = semantic alignment of matched pairs,
$q^{*}$ = actionability (discounted 80\% when no comments are confirmed),
$e$ = efficiency (confirmed / total agent comments), $h$ = hallucination rate
(FABRICATED fraction), $\rho$ = redundancy penalty, and $\phi$ = excess plausible
penalty above threshold $\tau{=}0.70$.
The aggregate benchmark score is the difficulty-weighted mean:
$S = \sum_i s_i w_i / \sum_i w_i$ where $w_i = \log(\text{human\_comments}_i + 1)$.

\subsection{Rubric Validation ($\kappa = 0.75$)}
\label{subsec:rubric}

We validate that the judge operationalises the published rubric by sampling 30
judge classifications (10 per category, stratified) and applying the rubric
independently using a blind template that omits the judge's verdict. Cohen's Kappa
between judge labels and rubric-derived labels: $\kappa{=}0.75$ (substantial
agreement, $N{=}30$, exact agreement 83.3\%).

\subsection{Cross-Judge Validation}
\label{subsec:crossjudge}

To confirm the A$>$B$>$C ordering is not a judge artefact, we reran the judge step
on a stratified 10-PR sample (\texttt{batch\_03}, GPT-4o-mini agent outputs,
$N{=}127$ comments) using Claude Sonnet~4.6. Overall inter-judge agreement:
78.7\% ($\kappa{=}0.616$). Both judges independently produce the same A$>$B$>$C
ordering (Table~\ref{tab:crossjudge}).

\begin{table}[H]
\centering
\caption{Cross-judge agreement by context configuration ($N{=}127$ comments,
  10-PR sample, GPT-4o-mini agent). Both judges reproduce A$>$B$>$C.}
\label{tab:crossjudge}
\small
\begin{tabular}{lcccc}
\toprule
\textbf{Config} & \textbf{$N$} & \textbf{Agreement}
  & \textbf{GPT-5.2 proxy} & \textbf{Sonnet proxy} \\
\midrule
\configA & 46 & 78.3\% & $-$0.035 & $-$0.033 \\
\configB & 40 & 72.5\% & $-$0.063 & $-$0.045 \\
\configC & 41 & 85.4\% & $-$0.066 & $-$0.070 \\
\midrule
Overall  & 127 & 78.7\% ($\kappa{=}0.616$)
  & \multicolumn{2}{c}{direction: A$>$B$>$C confirmed} \\
\bottomrule
\end{tabular}
\end{table}

The lowest agreement is at \configB\ (72.5\%), consistent with the Type2\_Contextual
collapse: agent comments generated with file content are more specific and therefore
harder to classify unambiguously. The highest agreement is at \configC\ (85.4\%),
where increased hallucination makes FABRICATED classifications more clear-cut.


\section{Baseline Evaluation on \bench}
\label{sec:results}

\subsection{Experimental Setup}
\label{subsec:setup}

We evaluate on a 100-PR stratified sample: Type1\_Direct (40 PRs), Type2\_Contextual
(40), and Type3\_Latent (20). Type2 and Type3 are deliberately over-sampled relative
to their 21\% and 13\% dataset prevalence to ensure statistical power for per-difficulty
analysis; per-difficulty scores are therefore reported separately and the weighted
aggregate $\bar{s}$ uses the actual 100-PR counts.

\subsection{Leaderboard Results}
\label{subsec:main}

\begin{figure}[H]
\centering
\includegraphics[width=0.90\linewidth]{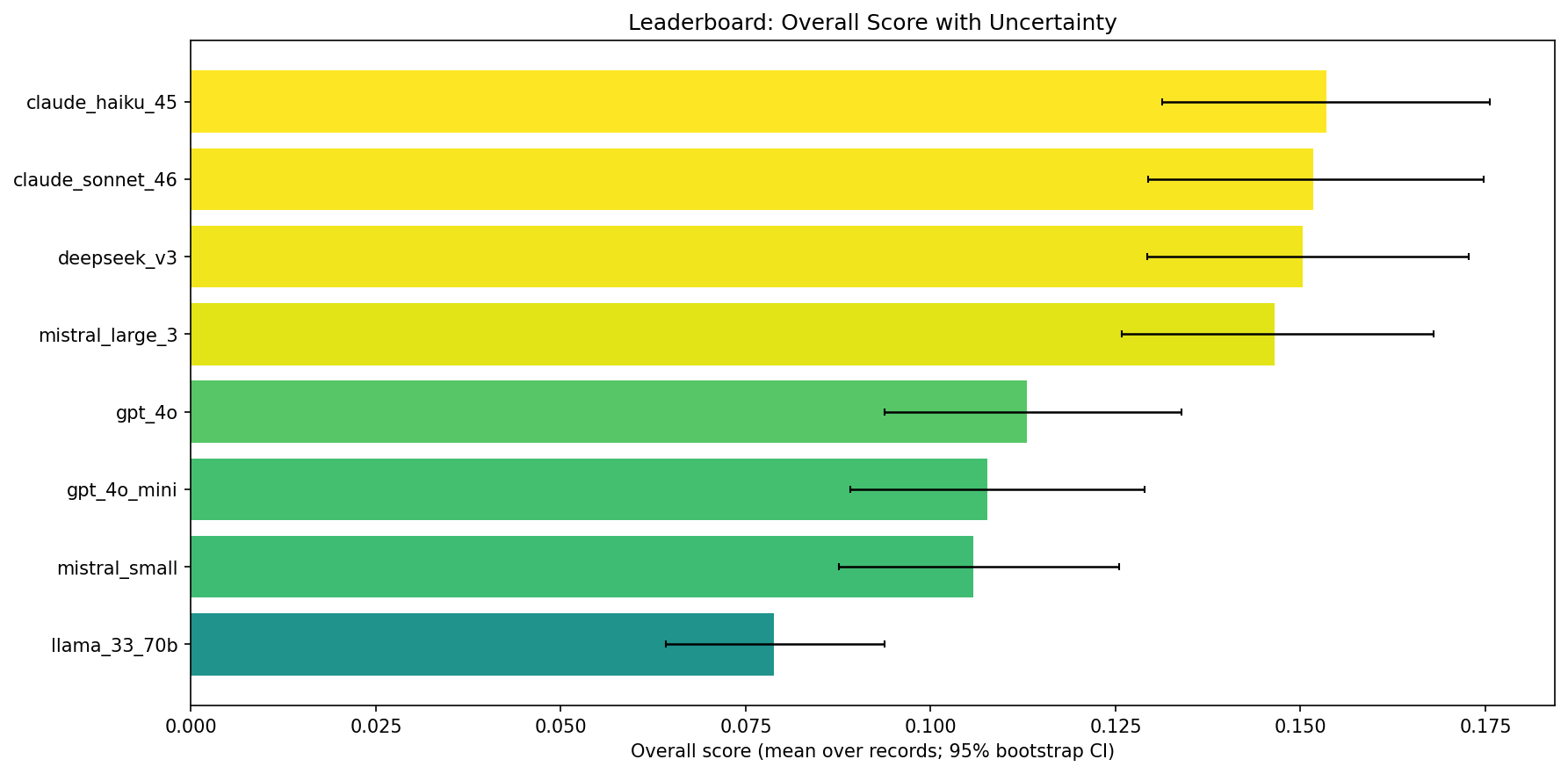}
\caption{\bench\ leaderboard with 95\% bootstrap CIs. Top four models cluster
  ($\bar{s}$~0.147--0.153); clear tier gap at rank~4/5.}
\label{fig:leaderboard_bar}
\end{figure}

Table~\ref{tab:leaderboard} establishes that \bench\ is non-trivial: no model detects
more than 31\% of human-flagged issues on any configuration, and the mean detection
rate across all 8 models on \configA\ is approximately 26\%. These results quantify
the gap between current AI code review capability and human expert performance.
Figure~\ref{fig:leaderboard_bar} shows 95\% bootstrap confidence intervals
($N{=}300$, $B{=}10{,}000$). The top four models
(Haiku, Sonnet, DeepSeek, Mistral Large; $\bar{s}$~0.147--0.153) are not
statistically distinguishable, while the rank-4/rank-5 gap (0.147 vs.\ 0.113) is
statistically significant.

\begin{table}[H]
\centering
\caption{\bench\ leaderboard: 100-PR sample, 8 models, sorted by mean
  difficulty-weighted composite $\bar{s}$ across all configurations.
  Parse failures in footnotes. Best per column in bold.}
\label{tab:leaderboard}
\small
\begin{tabular}{lcccccc}
\toprule
\textbf{Model} & \textbf{$s_A$} & \textbf{$s_B$} & \textbf{$s_C$}
  & \textbf{$\bar{s}$}
  & \textbf{DR\textsubscript{A}} & \textbf{FPR\textsubscript{avg}} \\
\midrule
Claude Haiku 4.5$^{\ddagger}$
  & 0.172 & 0.129 & 0.126 & \textbf{0.153}
  & \textbf{0.306} & 0.346 \\
Claude Sonnet 4.6$^{\dagger}$
  & \textbf{0.190} & 0.135 & 0.122 & 0.152
  & 0.297 & 0.227 \\
DeepSeek V3
  & 0.181 & 0.132 & 0.118 & 0.150
  & 0.312 & 0.315 \\
Mistral Large 3
  & 0.170 & 0.125 & 0.135 & 0.147
  & 0.305 & 0.353 \\
GPT-4o
  & 0.134 & 0.110 & 0.090 & 0.113
  & 0.220 & \textbf{0.193} \\
GPT-4o-mini
  & 0.110 & 0.095 & 0.093 & 0.108
  & 0.210 & 0.353 \\
Mistral Small
  & 0.131 & 0.080 & 0.091 & 0.106
  & 0.257 & 0.251 \\
Llama 3.3 70B$^{\S}$
  & 0.088 & 0.071 & 0.065 & 0.079
  & 0.223 & 0.417 \\
\bottomrule
\multicolumn{7}{l}{$^{\dagger}$~1 parse failure (hard-zeroed).
  $^{\ddagger}$~11 parse failures (hard-zeroed; removing them raises
  $\bar{s}$ by $\approx$0.006, no rank change).} \\
\multicolumn{7}{l}{$^{\S}$~30 judge parse fallbacks (score $\times 0.5$).} \\
\end{tabular}
\end{table}

DeepSeek~V3 achieves the highest raw detection on \configA\ (DR~=~0.312) and
outperforms all OpenAI models at approximately \$0.28/M input tokens versus \$2.50
for GPT-4o~\cite{openai2024pricing}: Tier~1 performance at roughly 9$\times$ lower
cost. GPT-4o achieves the lowest hallucination rate (FPR~=~0.193) but its detection
rate lags all upper-tier competitors.

\subsection{Context Configuration Behaviour}
\label{subsec:context_gating}

\begin{figure}[H]
\centering
\includegraphics[width=0.95\linewidth]{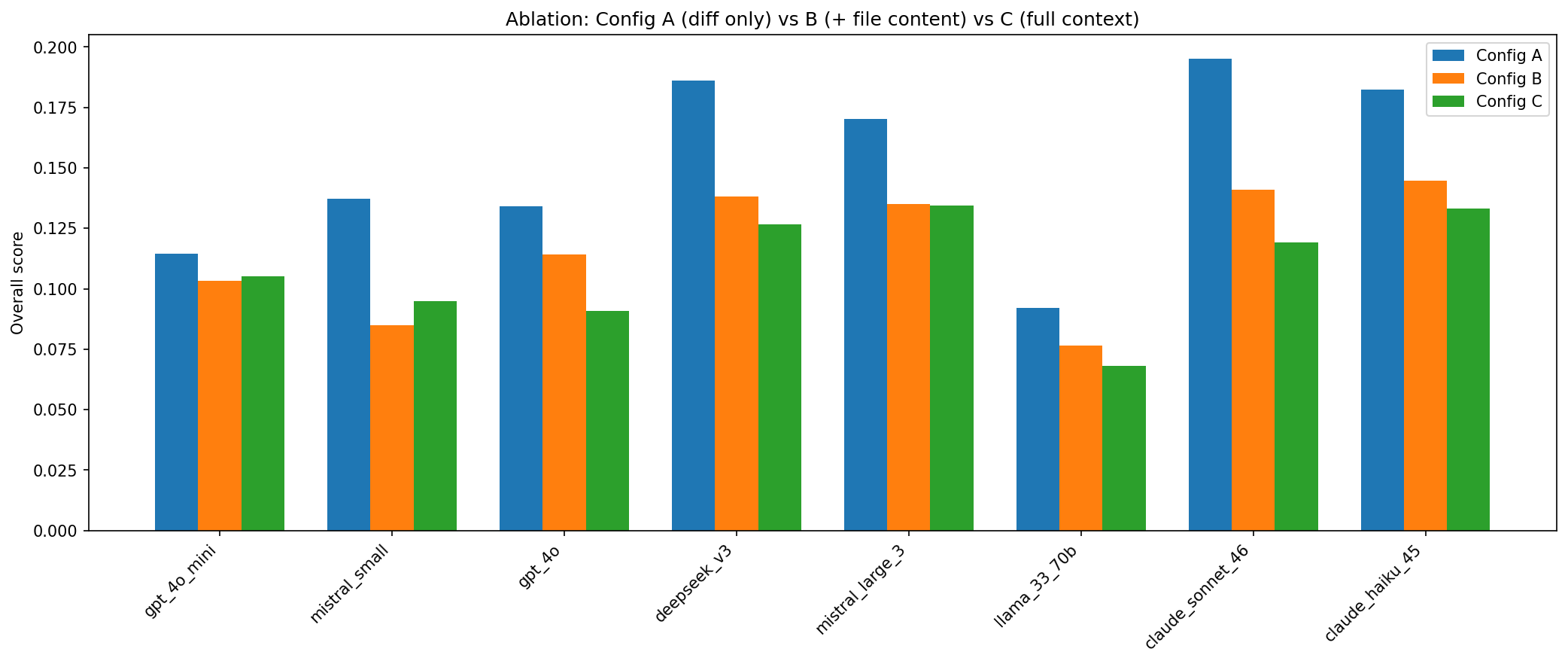}
\caption{Composite score by model and context configuration. All 8 models
  degrade A$\to$C; the A$\to$B drop is the largest step.}
\label{fig:config_ablation}
\end{figure}

All 8 models degrade monotonically from \configA\ to \configC\
(Figure~\ref{fig:config_ablation}). Critically, the three configurations differ in
\emph{layer composition}, not in token volume: \configA\ and \configC\ differ by
only 500 tokens (2,000 vs.\ 2,500). The A$>$B$>$C degradation therefore cannot be
attributed to context length alone. Rather, it implicates the specific content
added: execution context and behaviour mapping (\configB) and test signatures
(\configC) each introduce additional signal that models cannot reliably integrate,
even within a modest token budget. The V2 context builder uses AST-based function
extraction and import graph resolution, ruling out content selection as the cause.
The failure is in \emph{attention representation}: once relevant context is provided
as a flat token sequence alongside the diff, models cannot reliably direct attention
to the changed lines. The Type2\_Contextual collapse is the primary diagnostic:
Sonnet drops from 0.22 to 0.10 on contextual issues when execution context is added
at \configB, and DeepSeek drops from 0.20 to 0.10, consistent with the
attention dilution documented by Liu~et~al.~\cite{liu2024lost}. Table~\ref{tab:difficulty_ablation}
shows the full difficulty-by-configuration breakdown.

\begin{figure}[H]
\centering
\includegraphics[width=\linewidth]{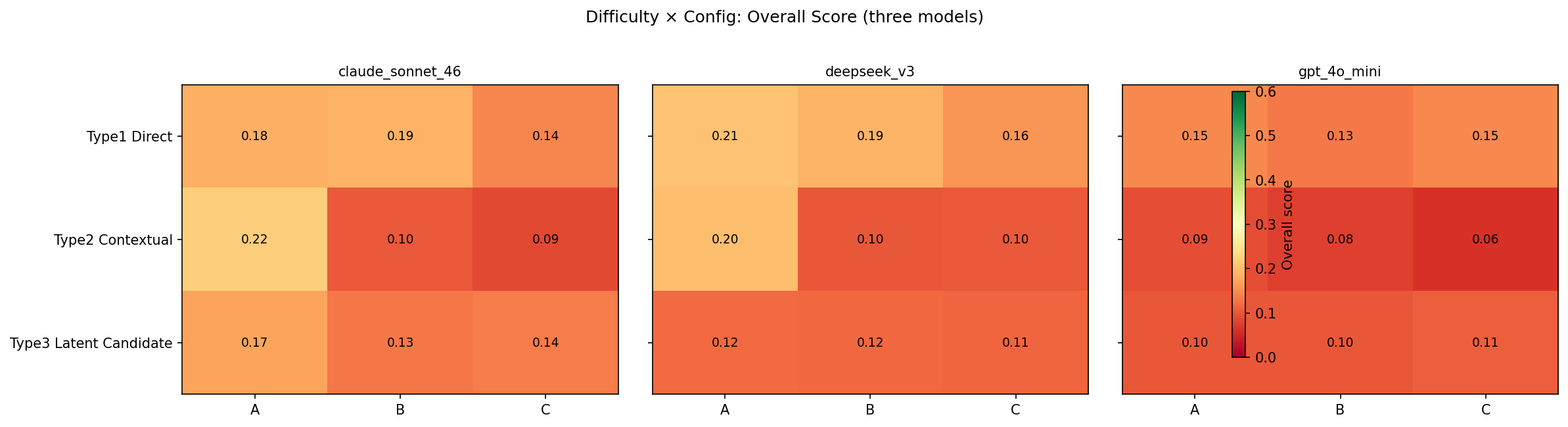}
\caption{Difficulty $\times$ config heatmap for three models. The Type2
  collapse at \configB\ is visible across all three.}
\label{fig:heatmap}
\end{figure}

\begin{table}[H]
\centering
\caption{Composite score by difficulty and config for three representative
  models. \textit{Collapse}: $>$50\% drop from \configA\ to \configB.}
\label{tab:difficulty_ablation}
\small
\begin{tabular}{llcccl}
\toprule
\textbf{Model} & \textbf{Difficulty} & \textbf{A} & \textbf{B} & \textbf{C}
  & \textbf{Trend} \\
\midrule
\multirow{3}{*}{Claude Sonnet 4.6}
  & Type1 & 0.18 & 0.19 & 0.14 & stable \\
  & Type2 & \textbf{0.22} & 0.10 & 0.09 & \textbf{collapse} \\
  & Type3 & 0.17 & 0.13 & 0.14 & stable \\
\midrule
\multirow{3}{*}{DeepSeek V3}
  & Type1 & 0.21 & 0.19 & 0.16 & gradual \\
  & Type2 & \textbf{0.20} & 0.10 & 0.10 & \textbf{collapse} \\
  & Type3 & 0.12 & 0.12 & 0.11 & stable \\
\midrule
\multirow{3}{*}{GPT-4o-mini}
  & Type1 & 0.15 & 0.13 & 0.15 & stable \\
  & Type2 & 0.09 & 0.08 & 0.06 & gradual \\
  & Type3 & 0.10 & 0.10 & 0.11 & stable \\
\bottomrule
\end{tabular}
\end{table}

\subsection{Hallucination Profiles}
\label{subsec:behaviour}

\begin{figure}[H]
\centering
\includegraphics[width=0.88\linewidth]{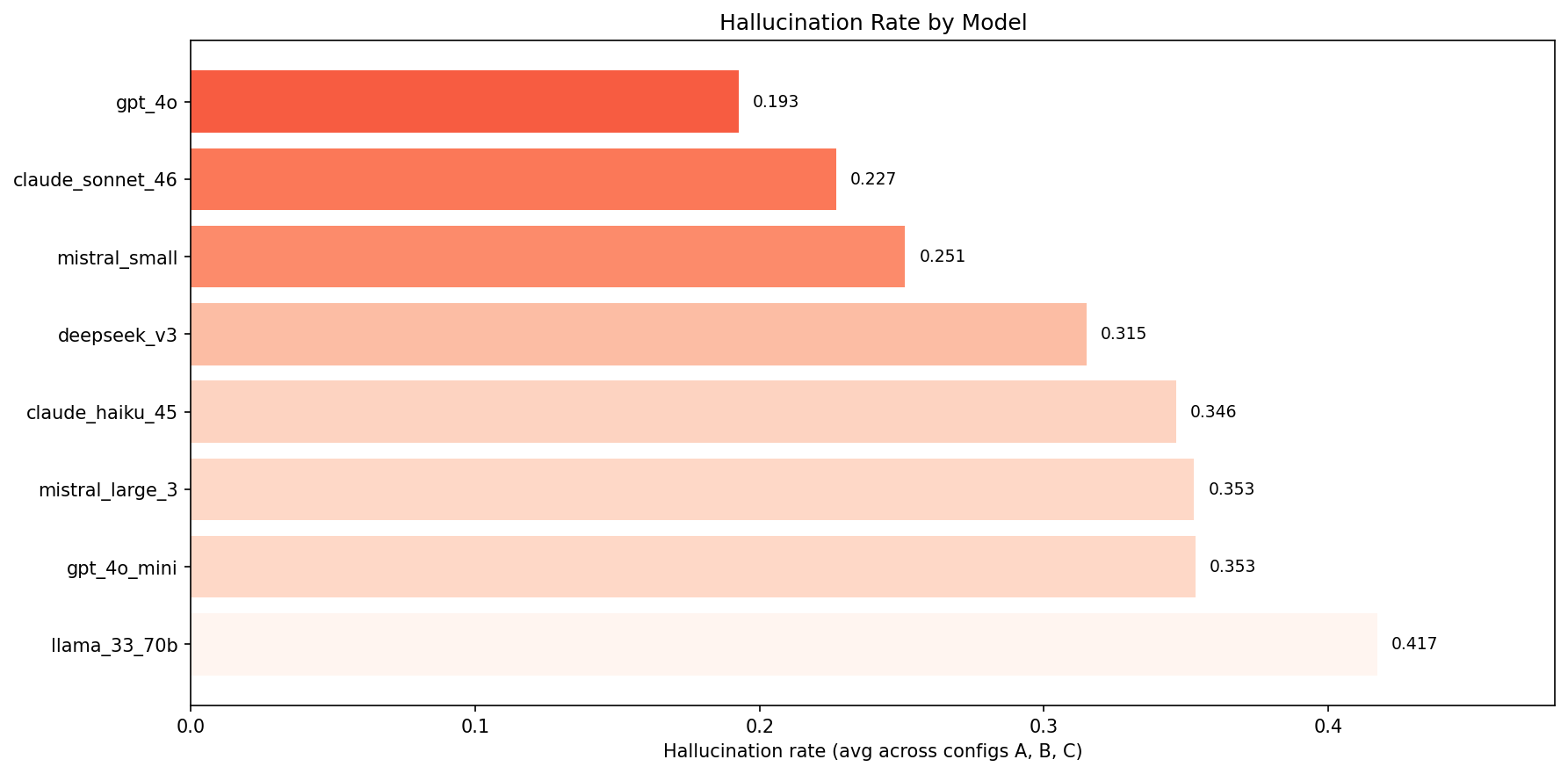}
\caption{Mean hallucination rate (FPR) across \configA/B/C.
  GPT-4o is most precise (0.193); Llama fabricates at 0.417.}
\label{fig:hallucination}
\end{figure}

Figure~\ref{fig:hallucination} shows a clear precision-recall tradeoff: Haiku and
DeepSeek achieve higher detection but at higher hallucination cost; Sonnet and GPT-4o
produce more reliable comments with lower FPR. The composite scorer correctly
captures this: Sonnet ranks second despite lower raw DR than Haiku, because the
hallucination and redundancy penalties offset Haiku's recall advantage.
Batch ordering is stable across all 10 batches, confirming the ranking is not
a sample artefact.

\textbf{Practical deployment guidelines.} (1)~Use diff-only context (\configA):
adding file content does not help and actively harms performance across all tested
models. (2)~Prioritise models by hallucination rate: GPT-4o (FPR~=~0.193) and
Sonnet (FPR~=~0.227) produce more reliable comments than higher-recall alternatives,
reducing triage burden. (3)~DeepSeek~V3 provides Tier~1 detection at roughly
9$\times$ lower API cost than GPT-4o with acceptable precision.

\subsection{Benchmark Properties}
\label{subsec:rq_summary}

\begin{table}[H]
\centering
\caption{Summary of benchmark properties and research question answers.}
\label{tab:rq}
\small
\begin{tabular}{clp{5cm}l}
\toprule
\textbf{RQ} & \textbf{Question} & \textbf{Answer} & \textbf{Evidence} \\
\midrule
1 & Detection rate and FPR & 15--31\% on \configA; FPR 0.193--0.417 & Table~\ref{tab:leaderboard} \\
2 & Context direction & All models degrade A$>$B$>$C & Figure~\ref{fig:config_ablation} \\
3 & Uniformity & Universal direction; Type2 drives the A$\to$B drop & Table~\ref{tab:difficulty_ablation} \\
4 & Hard categories & Type2 collapses at \configB; Type3 near-zero & Figure~\ref{fig:heatmap} \\
\bottomrule
\end{tabular}
\end{table}


\section{Dataset Analysis}
\label{sec:dataanalysis}

Table~\ref{tab:difficulty_language} shows the full difficulty-by-language
distribution. Python has the highest absolute Type3 count (32 of 43, 74.4\%),
consistent with Python's dynamic import patterns. Go and TypeScript have the
highest proportional Type2 rates (22.9\% and 33.3\%), consistent with interface
contracts and type propagation creating contextual issues. PR size spans 5--797
added lines (mean 218.2) and 1--83 files changed (mean 9.2), covering both
focused patches and broad refactors.

A \bench\ task is hard in three structurally distinct ways: Type1 tasks require
identifying a fault in changed lines without hallucinating non-existent bugs; Type2
tasks require understanding changed code in relation to surrounding context, but the
current flat-token representation makes this harder with more context; Type3 tasks
require cross-file dependency reasoning from import structure alone. The AI ceiling
is clear: the best model (Haiku, DR$_A$~=~0.306) detects roughly one in three
direct issues. If a second human reviewer achieves 50--70\% recall~\cite{bacchelli2013expectations},
the gap to human performance is approximately 20--40 percentage points.

\section{Limitations}
\label{sec:limitations}

\textbf{Scope of the context degradation finding.} The V2 context builder uses
AST extraction and import graph resolution, not raw file dumping. The degradation
therefore implicates attention representation rather than content selection. Context
provision strategies encoding relevance at the token level, such as explicit
changed-vs-unchanged boundary markers or interleaved annotations, may produce
different outcomes and are a direct priority for future work.

\textbf{Language and sample scope.} The dataset is Python-dominant (69.1\%) and
evaluated on a 100-PR sample from the 350-PR full corpus. Results for non-Python
languages and per-difficulty sub-group statistics should be interpreted with this
in mind. Full 350-PR evaluation is deferred to the camera-ready version.

\textbf{Evaluation methodology.} The primary judge is GPT-5.2. Cross-judge
validation confirms the A$>$B$>$C direction under Sonnet~4.6 ($\kappa{=}0.616$),
but absolute score values differ between judges. The potential for mild judge-family
bias cannot be excluded without cross-family validation on the full leaderboard.
No human performance baseline is available; the 20--40 percentage-point gap estimate
to human recall is derived from prior literature~\cite{bacchelli2013expectations}.

\textbf{Contamination.} Despite recency filtering, RQS contamination weighting,
and GPL preference, we cannot fully exclude the possibility that some benchmark PRs
appeared in model pretraining. Type3 issues require cross-file reasoning that cannot
be recalled from a memorised patch, providing partial contamination resistance for
the hardest difficulty tier.


\section{Conclusion}
\label{sec:conclusion}

We introduced \bench, a benchmark of 350 pull requests with human-annotated ground
truth for evaluating AI code review quality. The dataset, pre-built context
artifacts and evaluation harness are released publicly:

\begin{itemize}
  \item \textbf{Dataset:} \url{https://huggingface.co/datasets/foundry-ai/swe-prbench}
  \item \textbf{Evaluation harness:} \url{https://github.com/FoundryHQ-AI/swe-prbench}
\end{itemize}

Baseline results on 8 frontier models establish that no model detects more than 31\%
of human-flagged issues on any configuration, confirming that AI code review remains
far from human expert performance. The context configuration ablation reveals that
even structured semantic context --- built via AST extraction, import graph
resolution, and behaviour mapping --- degrades performance relative to diff-only
prompts. This implicates attention representation rather than content selection as
the binding constraint: models cannot reliably distinguish changed from unchanged
lines once both appear in a flat token sequence. We invite the community to submit
results to the \bench\ leaderboard, extend the dataset to additional languages and
domains, and develop context representation strategies --- such as explicit
changed-vs-unchanged boundary markers or retrieval-augmented context --- that close
the gap between current AI performance and human expert review quality.

\medskip
\noindent\textbf{Reproducibility Note.}
Scores reported in this paper reflect pipeline version \texttt{v0.4.1} with
GPT-5.2 as judge at temperature~$= 0$. Frontier model APIs do not guarantee full
determinism at temperature~$= 0$, so minor score variation across independent runs
is expected. The two-tier ranking structure and A$>$B$>$C ordering are stable
across runs and confirmed by cross-judge validation ($\kappa = 0.616$,
Section~\ref{subsec:crossjudge}).

\bibliographystyle{plainnat}


\appendix

\section{PR Record Schema}
\label{app:schema}

Each \bench\ pull request is stored as a JSON record with five field categories.
\textbf{Metadata:} \texttt{task\_id}, \texttt{repo}, \texttt{pr\_number},
\texttt{language}, \texttt{pr\_type}, \texttt{difficulty}, \texttt{merged\_at}.
\textbf{Scoring signals:} \texttt{rvs\_score} and five component scores
(\texttt{review\_depth}, \texttt{code\_complexity}, \texttt{discussion\_signal},
\texttt{test\_change\_signal}, \texttt{bug\_fix\_signal}), plus
\texttt{lines\_added}, \texttt{lines\_removed}, \texttt{files\_changed}.
\textbf{Reproducibility anchors:} \texttt{base\_commit} and \texttt{head\_commit}
SHA hashes pinning the exact file state at review time.
\textbf{Ground truth:} \texttt{human\_review\_comments} array with per-comment
fields for author, body, path, line, diff hunk, and reply status; plus
\texttt{num\_substantive\_comments}, \texttt{num\_unique\_reviewers},
\texttt{has\_requested\_changes}, \texttt{ai\_comments\_removed}.
\textbf{Diff:} \texttt{diff\_patch} as the raw unified diff retrieved from the
GitHub REST API; \texttt{agent\_input} is populated by the context builder at
evaluation time.

A representative record (dask/dask PR~\#12221) is available in the HuggingFace
dataset card. This PR carries RVS~=~0.334, placing it below the 0.35 inclusion
threshold; it is provided as a boundary illustration of the quality cut, not as a
dataset member.

\section{Context Builder Layer Specification}
\label{app:layers}

The V2 context builder assembles each configuration through a layered pipeline with
budget rollover (unused tokens flow to subsequent layers). Layer 1a (metadata,
200 tokens fixed): PR title, repository, difficulty type, language, file count,
truncated description; never includes RVS score, ground truth, or post-merge data.
Layer 1b (key changes summary, 200 tokens): LLM-generated natural-language summary
of principal changes inserted before the diff. Layer 2 (diff, 800--1,200 tokens):
per-file hunks sorted by importance $= 0.5 \times \text{lines\_changed} + 0.3 \times
\text{comment\_density} + 0.2 \times \text{function\_changes}$; production files
before test files; at most 4 files. Layer 3 (changed files, 400--800 tokens): AST
extraction for Python (\texttt{get\_innermost\_enclosing\_nodes} plus relevant
imports), cluster-based line windows for other languages; always uses
\texttt{base\_commit}, never \texttt{head\_commit}; excludes test files. Layer 4a
(related files, 100--300 tokens): import graph resolution for Python; GitHub code
search for others; empty with \texttt{not\_applicable\_type1} flag for Type1 PRs.
Layer 4b (test files, 150--300 tokens): body-stripped, retaining \texttt{def},
\texttt{async def}, \texttt{class}, \texttt{assert}, and \texttt{@fixture} lines
only; Go/TypeScript PRs also receive a behaviour mapping layer (100--150 tokens)
making config-to-implementation propagation chains explicit.

\section{Evaluation Protocol Detail}
\label{app:protocol}

\textbf{CONFIRMED criteria (all must hold):} (1) the agent identifies a specific
issue in the code; (2) a human reviewer identified the same underlying issue; (3)
the issue concerns the same file or functional area; (4) the agent concern would
lead to the same kind of code change as the human concern.
\textbf{PLAUSIBLE:} grounded in visible code, factually correct, reasonable
engineering concern, but not matched to any human comment.
\textbf{FABRICATED (any sufficient):} references code not in context; makes
factually incorrect claims; describes a non-existent bug; invents method signatures
or variable names.

\textbf{Matching detail.} Similarity uses \texttt{all-mpnet-base-v2} sentence
embeddings. Stage 1: judge-CONFIRMED pairs are sorted by combined
actionability-and-similarity signal; the top-scoring 1-to-1 assignment is selected
greedily. Stage 2: unmatched non-FABRICATED comments are evaluated against
remaining human comments via pairwise affinity $= \text{cosine}(e_a, e_h) + 0.15
\cdot \mathbf{1}[\text{same file}] + \delta_{\text{line}}$, clamped to $[0,1]$.
Pairs below 0.30 are discarded. The threshold was calibrated against the 30-sample
rubric validation set; values below 0.30 produced no judge-CONFIRMED pairs that
also passed blind human review. Agent comments unmatched after both stages and
not FABRICATED are reclassified to PLAUSIBLE.

\end{document}